\documentclass[aps,prd,preprint,nofootinbib,groupedaddress]{revtex4-1}

\usepackage[latin1]{inputenc}
\usepackage{slashed}
\usepackage{amssymb} 
\usepackage{amsmath}
\usepackage{multirow}
\usepackage{graphicx}
\usepackage{hyperref}

\makeatletter 
\def\p@subsection{}
\def\p@subsubsection{}
\makeatother

\newcommand{\eps}{\varepsilon}
\newcommand{\beq}{\begin{eqnarray}}
\newcommand{\eeq}{\end{eqnarray}}
\newcommand{\nn}{\nonumber}

\newcommand{\hideit}[1]{ }

\newcommand{\system}[1]{\left\{\begin{array}{l} #1\end{array}\right.}

\newcommand{\Kahler}{K\"ahler }

\newcommand{\vev}[1]{\langle #1\rangle}

\newcommand{\sut}{$SU(3)$\,}
\newcommand{\uone}{$U(1)$\,}
\newcommand{\rps}{$\slashed{R}_p$}
\newcommand{\rpv}{$R$-parity violation}
\newcommand{\supo}{superpotential}

\newcommand{\hors}{horizontal symmetry}
\newcommand{\nbmu}{n_{\bar\mu}}
\newcommand{\nmu}{n_{\mu}}
\DeclareMathOperator{\diag}{diag}

\newcommand{\HH}{\mathcal{H}}
\newcommand{\OO}{\mathcal{O}}

\begin{document}


\preprint{arXiv:1305.2921 [hep-ph]}
\preprint{SCIPP 13/08}



\title{Natural, $R$-parity violating supersymmetry and\\ horizontal flavor symmetries}


\author{Angelo Monteux}\email{amonteux@ucsc.edu}
\affiliation{Santa Cruz Institute for Particle Physics and\\
 Department of Physics, 
  University of California, Santa Cruz CA 95064}



\begin{abstract}
The absence so far of any supersymmetric signals at the LHC pushes towards a rethinking of the  assumptions underlying the MSSM. Because the large missing $E_T$ searches are inadequate to detect a LSP decaying within the detector, $R$-parity violating supersymmetry is still a good candidate for low energy, natural supersymmetry. 
We show that, in Froggatt-Nielsen-like models of horizontal symmetries, specific textures for the $R$-parity violating couplings are dictated by the symmetry, with the largest coupling involving the third generation fields.
Lepton number can be an accidental symmetry of the renormalizable superpotential and baryon number violation is given by a $\bar u\bar d\bar d$ operator. The collider phenomenology then mimics the main features of  MFV R-parity violating supersymmetry.
The LSP can evade current LHC supersymmetry searches, is allowed to be well below 1 TeV and at the same time all the constraints from proton decay and other low energy decays can be satisfied; in particular, dimension five operators allowed by $R$-parity but dangerous for the proton are under control, while neutrino masses are generated by the Weinberg operator. 
Assuming sub TeV ({\it natural}) superpartners, we obtain both upper and lower limits on the magnitude of the dominant $R$-parity violating coupling: a lower limit of order $10^{-9}$ arises from null LHC searches on $R$-hadrons and heavy stable charged particles, while a upper limit of order $10^{-2}$ follows from constraints on low-energy flavor changing neutral currents. Displaced vertices are predicted in the lower end of this range.

\end{abstract}


\maketitle

\section{Introduction}

Generic supersymmetric extensions of the Standard Model do not share its success in suppressing large flavor changing effects: first, baryon and lepton number are not accidental symmetries of the Supersymmetric Standard Model, unless some additional (usually discrete) symmetry is assumed, e.g. $R$-parity, $R_p=(-1)^{2S+3B+L}$ \cite{Farrar:1978xj}, or matter parity $M_p$ \cite{Bento:1987mu}, under which the SM fields are even and the superpartners are odd. This way, all the baryon number violating (BNV) and lepton number violating (LNV) dimension four operators are forbidden; still, dimension five operators that  would induce proton decay are allowed by $R$-parity. Other possibilities are baryon triality ($B_3$) \cite{Ibanez:1991pr}, that just forbids baryon number violation, or Proton Hexality ($P_6$) \cite{Dreiner:2005rd}, the direct product of $M_p$ and $B_3$.

Second, even with a discrete symmetry, generic squark masses at the TeV scale would generate unsuppressed flavor-changing neutral currents (FCNCs), which  contribute to low-energy phenomena such as meson mixing and decays. To eliminate large FCNCs in low energy supersymmetry, squark degeneracy is usually assumed; then, in a GIM-like mechanism, their contribution is suppressed. Although natural (at the SUSY breaking scale) in  gauge mediated models of supersymmetry breaking, squark degeneracy is not guaranteed in other frameworks, such as gravity mediation. An alternative way to suppress large FCNC is to assume alignment between  quark and squark mass matrices: that is, assuming that the squark mass-squared matrices and the quark mass matrices are simultaneously diagonal, in the basis where the gluino interactions are diagonal as well \cite{Nir:1993mx}. Nevertheless, as pointed out in \cite{Blum:2009sk,Gedalia:2009kh}, for a SUSY breaking scale of $1$ TeV alignment alone is not enough to satisfy the constraints from both $K-\overline K$ and $D-\overline D$ mixing, and an $\OO(10\%)$ degeneracy for the squarks is needed.

A third, related problem involves the mass of the Higgs boson and the scale of supersymmetry. In the MSSM, the tree-level Higgs boson mass is bounded above by $M_Z$, and one has to rely on radiative correction to lift it up to the value $m_h=126$ GeV discovered at the LHC \cite{Aad:2012tfa,Chatrchyan:2012ufa}. This either implies a heavy stop or large soft $A$-terms, or both. On the other hand, the fine-tuning of the weak scale is also sensitive to the stop mass, and heavy stops lead to higher degrees of fine-tuning. The exact level depends on the definition of fine-tuning, and it is debated which degree of fine-tuning is acceptable and which is not, but it is generically accepted that stops heavier than 1 TeV are a problem for the naturalness paradigm (see ref. \cite{Feng:2013pwa} for a review on the concept of naturalness in light of the LHC searches). When considering light superpartners, a light Higgs at $126$ GeV requires extra contributions to its mass. Here, we will be assuming the presence of a NMSSM-like singlet $N$, with a superpotential term $\lambda N\phi_u\phi_d$; as shown in \cite{Hall:2011aa}, for $\lambda\sim0.7$ this allows sub-TeV stops and a fine-tuning of order $10\%$.

The minimal supersymmetric standard model (MSSM) usually just assumes $R$-parity, even though additional suppression is needed for the dimension five operators; in this scenario superpartners are always pair-produced. The lightest supersymmetric partner (LSP) is stable and usually a neutralino, thus providing a WIMP dark matter candidate. Any superparticle produced at the LHC decays in a cascade leading to the LSP, which escapes the detector generating events with large missing energy $\slashed E_T$. In most of the MSSM parameter space, LHC has set lower limits  of 1 TeV or above for superpartner masses, although searches directed specifically to the stop give slightly lower bounds, of about 650 GeV. Still, the high limits on the gluino mass result in fine-tuning of the weak scale as the gluino mass enters the RGEs of the squark masses, including the stop. This has pushed some to abandon the concept of naturalness and accept that some parameters of our theories might be fine-tuned and that their smallness might be due to environmental selection.

Nevertheless, many low-energy ({\it{natural}}) supersymmetric models exist that evade LHC searches: compressed SUSY \cite{Martin:2007gf}, stealth SUSY \cite{Fan:2011yu}, and several models of $R$-parity violating SUSY \cite{Barbier:2004ez,Allanach:2012vj,Evans:2012bf,Nikolidakis:2007fc,Csaki:2011ge,Bhattacherjee:2013gr}; the common aspect of this class of models is that the decay cascades have small missing energy, thus evading the requirement of large $\slashed E_T$ in ATLAS and CMS searches.

In this work, we will focus on the last possibility, \rpv; without $R$-parity, the LSP is not stable and it can decay within the detector, thus leaving no missing energy. Moreover, because the LSP does not need to be the dark matter, it can as well be a charged or colored particle (the gravitino can be a dark matter candidate, provided that its lifetime is long enough on cosmological scales \cite{Takayama:2000uz}). As we have seen, although $R$-parity was initially proposed to stabilize the proton, it is not enough for this purpose, and its presence at all can be questioned.
In this context, lepton and baryon number conservation is just approximate, and one can explain small violations as the result of a broken symmetry (a $SU(3)^5$ flavor symmetry group in Minimal Flavor Violation \cite{Nikolidakis:2007fc}, or a \uone\ in the models we will be considering).

A simple model of \rpv \, involves a horizontal symmetry $U(1)_\HH$ (which might also be an $R$-symmetry)  responsible for the hierarchy in the SM fermion masses and mixings, in a supersymmetric extension of the Froggatt-Nielsen mechanism \cite{Froggatt:1978nt,Leurer:1992wg,Nir:1993mx,Leurer:1993gy,BenHamo:1994bq}. The high-energy theory is assumed to be invariant under a horizontal symmetry, broken by the vacuum expectation value of a field $S$ with charge $\HH[S]=-1$ (the flavon). In the low-energy theory, heavy fields that have been integrated out generate effective operators proportional to the spurion $\eps={\vev{S}\over M}$, where $M$ is the high scale related to the horizontal symmetry breaking mechanism.\footnote{The high energy theory  generally includes extra charged fields, but we are not interested in its specific form; for a horizontal \uone, heavy mirror fermions are integrated out at the scale $M$ \cite{Froggatt:1978nt,Leurer:1993gy}. UV models for MFV SUSY were proposed in \cite{Krnjaic:2012aj,Franceschini:2013ne,Csaki:2013we}, while a model in which the \rps\ couplings arise through SUSY-breaking soft terms was studied in \cite{Krnjaic:2013eta}.
}
 Only terms that are invariant under the symmetry are allowed in the superpotential.  In order to give mass to the SM fermions, the Higgs Yukawa couplings must be allowed, and their hierarchies appear because they are proportional to different powers of $\eps$, corresponding to the diverse horizontal charges of the SM fields. For the same reason, the operators that break $R$-parity can be small \cite{BenHamo:1994bq}, while  FCNCs can be suppressed by charge assignments that give either squark degeneracy or quark-squark  alignment.

Here we will show that the horizontal symmetry  predicts the relative hierarchy of the the $R$-parity violating couplings  and proton decay can be sufficiently suppressed. The horizontal symmetry can also be embedded in the flavor group $U(3)^5$ that is a symmetry of the Lagrangian in the limit of vanishing Yukawa couplings, in a weaker version of the Minimal Flavor Violation hypothesis. Because no continuous global symmetry is expected in quantum gravity, the horizontal symmetry might be a discrete symmetry $Z_N^\HH$: then, $S^N$ would have horizontal charge $0\,\text{mod}N$ and the maximal suppression for operators would be $\eps^{N-1}$; achieving adequate suppression for $R$-parity violating decays would push $N$ to be greater than 10, making the model less attractive; if the horizontal symmetry is $Z_{N_1}^\HH\times Z_{N_2}^\HH$, with two spurions $S_1$ and $S_2$, the values of $N_1$ and $N_2$ can be lower.

We will not ask for the \hors\, to be anomaly-free, or that its anomalies with respect to the SM gauge groups are universal, because we do not want to commit to a specific high-energy model; as argued in \cite{Dine:2012mf}, anomaly universality is specific to the Green-Schwartz mechanism in the heterotic string, involving one dilaton field; for both continuous and discrete symmetries, there are examples in the heterotic \cite{Ludeling:2012cu} and type II \cite{Ibanez:1998qp,Dine:2004dk} string theory where the anomalies are cancelled by multiple moduli which do not couple universally to gauge field strengths of different gauge groups. Furthermore, we are considering an effective theory where  heavy charged fields have been integrated out at a high scale where the symmetry breaks down, so additional heavy states would contribute to the anomalies.  We conclude that the only constraints on the \hors \, are given by the quark and lepton masses and mixings. Since there are more variables than constraints, some of the horizontal charges are free parameters.

While this paper was in preparation, a similar work  was published \cite{Florez:2013mxa}; the authors of ref. \cite{Florez:2013mxa} also study $R$-parity violating SUSY with an horizontal symmetry, and in particular how baryonic \rpv\ can evade the stringent LHC supersymmetry searches. A standard Green-Schwarz mechanism with universal anomalies was assumed. As stated above, we do not ask for anomaly universality, and clarify that the structure of the RPV coefficients is univoquely determined by the masses and mixings of the SM fermions. An important difference arising from anomaly universality involves the $\mu$ term: as can be seen in ref. \cite{Dreiner:2003hw}, anomaly cancellation through a Green-Schwartz mechanism  gives a specific prediction for the magnitude of the $\mu$ term, $\mu=m\eps^{|\nmu|}, \eps^{\nmu}=\frac{m_d m_sm_b}{m_em_\mu m_\tau}=5.14^{+4.40}_{-3.02}$, where the error bars come from the experimental errors in the quark masses, $\eps=0.226$ and $m=M_P$ if $\nmu \geq 0$, or $m=m_{3/2}$ for negative $\nmu$. This corresponds to $\nmu=0$ or $-1$ or$-2$, with $-1$ the preferred value: this is a solution of the $\mu$ problem if the SUSY scale is not very much above 1 TeV. Although the main purpose of the present work is to consider low energy supersymmetry, a horizontal symmetry with non-universal anomalies leaves the possibility of  a hierarchy between the SUSY breaking scale and the weak scale. The only phenomenological requirement is $\nmu<0$, and $|\nmu|$ can easily be a number of order 10, thus explaining a hierarchy of up to $10^{-7}$ between $m_{3/2}$ and $\mu$. In this sense, it is noteworthy that, if there are no superpartners up to a few TeV, a horizontal symmetry with  cancellation of universal anomalies (as the one considered in \cite{Florez:2013mxa}) would be ruled out.

In addition, the present work takes a different approach to low energy supersymmetry, as it is pointed out that current LHC searches also exclude models with arbitrarly small \rps\ coefficients, and a range of allowed $R$-parity violating couplings is provided. Finally, we solve issues arising from low energy supersymmetry, as the  high value of the Higgs mass and the absence of FCNCs.

This paper is organized as follows: in section \ref{masses}, we will review how a  \hors\,  can generate hierarchies in the SM spectrum, and the phenomenological constraints on the charges of the fields. We will then investigate the implications for the \rps\   couplings in the superpotential in section \ref{rpv}, and discuss the phenomenological implications for both low-energy flavor physics and LHC signatures; requiring low energy supersymmetry will provide a lower bound for the \rps\ coefficients. In section \ref{align}, we will consider quark-squark alignment and whether light superpartners are compatible with low-energy flavor physics: this way, we reconcile the demand for low-energy SUSY from naturalness and the demand of a higher scale of supersymmetry from the absence of FCNC signals. In section \ref{NMSSM}, we examine the NMSSM contribution to the Higgs mass and present  horizontal symmetry constraints on that sector. In section \ref{mfv}, we will investigate the origin of the horizontal symmetry by asking it to be a subgroup of the $SU(3)^5$ flavor group; this corresponds to use a flavor symmetry that is already manifest in the SM; we will discuss the similarities of our approach to the MFV approach, and the different phenomenological implications.

\section{Horizontal Symmetry}
\label{masses}

In this section, following refs. \cite{Leurer:1992wg,Dreiner:2003hw}, we  construct an effective theory in which a \hors\, $\HH$ is responsible for the hierarchies and mixings of the SM fermion sector. Unlike refs. \cite{Dreiner:2003hw,Sierra:2009zq}, we do not assume anomaly cancellation through a GS mechanism. The horizontal symmetry is broken when a field $S$ with charge $-1$ acquires a vev $\vev{S}$ and the effective theory is valid up to the scale $M$, where the flavon dynamics takes place.\footnote{The flavor physics scale is unconstrained; it could be as low as $10^{3-4}$ TeV \cite{Leurer:1993gy}, or up to $M_P$  if string theory is responsible for it \cite{Ibanez:1994ig}.}
The MSSM superpotential is replaced by an expression that preserves $\HH$:
\beq\label{supoY}
\mu\phi_u\phi_d+Y^d_{ij}\phi_dQ_i\bar d_j+Y^u_{ij}\phi_uQ_i\bar u_j+Y^\ell_{ij}\phi_dL_i\bar \ell_j\,,\qquad i,j=1,2,3\ ,
\eeq
becomes 
\begin{align}
m \left(\frac{\vev S}M\right)^{|n_\mu|}\phi_u\phi_d+
\left(\frac{\vev S}M\right)^{m_{ij}} \phi_dQ_i\bar d_j+\left(\frac{\vev S}M\right)^{n_{ij}}\phi_uQ_i\bar u_j+\left(\frac{\vev S}M\right)^{p_{ij}}\phi_dL_i\bar \ell_j=\nn\\
=m\eps^{|n_\mu|}\phi_u\phi_d+\eps^{m_{ij}} \phi_dQ_i\bar d_j+\eps^{n_{ij}}\phi_uQ_i\bar u_j+\eps^{p_{ij}} \phi_dL_i\bar \ell_j\ ,\label{mssm}
\end{align}
where we have neglected $\OO(1)$ coefficients of the effective operators in the last equation and defined $\eps=\vev S/M$. If the exponents $m_{ij}, n_{ij}$ and $p_{ij}$ are {\it non-negative} (since the \supo\, is holomorphic in $S$) and {\it non-fractional} (unless the effective operator arises from some non-perturbative effect) the corresponding operator is allowed, otherwise it is forbidden. Below the SUSY breaking scale the potential does not have to be holomorphic, and operators with negative powers of $\eps$ appear from \Kahler corrections (for a complete discussion, see \cite{Dreiner:2003hw}): for dimensionless couplings, this generates an additional suppression of order $m_{3/2}/M_P$ and we will not consider it, while a $\mu$ term $\mu=m_{3/2}\eps^{-\nmu}$ is generated by a \Kahler correction $X\phi_u\phi_d\left(\frac{S^*}M\right)^{-n_\mu}$ in a Giudice-Masiero-like mechanism \cite{Nir:1995bu}, if $n_\mu=\HH[\phi_u]+\HH[\phi_d]-r<0$. In the expression above, we inserted a mass scale $m$ that is $M$ if $n_\mu$ is positive and is $m_{3/2}$ if $n_\mu<0$. For negative $\nmu$, the $\mu$ term is automatically suppressed with respect to  $m_{3/2}$.

Because $\HH[S]=-1$, the exponents are
\beq
&&m_{ij}=\HH[\phi_d]+\HH[Q_i]+\HH[\bar d_j]-r;\nn\\
&&n_{ij}=\HH[\phi_u]+\HH[Q_i]+\HH[\bar u_j]-r;\\
&&p_{ij}=\HH[\phi_d]+\HH[L_i]+\HH[\bar \ell_j]-r,\nn
\eeq
where $r=0,2$ takes care of the possibility that the horizontal symmetry is an $R$-symmetry. 
In the following we will denote the horizontal charge of a field $\Phi_i$ by the symbol of the field itself, $\HH[\Phi_i]=\Phi_i$, and the intergenerational difference between charges as $\HH[\Phi_i]-\HH[\Phi_j]=\Phi_{ij},\ i,j=1,2,3$.

\paragraph{Quark sector:}
Given the superpotential \eqref{mssm}, the masses and mixing angles in the quark sector can be expressed in terms of $\eps$ \cite{Froggatt:1978nt}: 
\beq
Y^a_{ij}&\sim& v_a \eps^{\phi_a+Q_i+a_j-r}\qquad a=u,d;\ i,j=1,2,3; \\
 \frac{m^a_i}{m^a_j}&\sim&\eps^{Q_i+a_i-Q_j-a_j},  \quad  |V_{ij}|\sim\eps^{|Q_i-Q_j|}, \nn
\eeq
where $V$ is the CKM  matrix. It can be written in the Wolfenstein parametrization,
\beq\label{ckm}
|V|=\left|\left(\begin{array}{ccc}1-\eps^2/2 &\eps&A\eps^3(\rho-i\eta)\\
-\eps &1-\eps^2/2 &A\eps^2 \\
A\eps^3(1-\rho-i\eta)&-A\eps^2&1
\end{array}\right)\right|
\sim
\left(\begin{array}{ccc}1 & \eps & \eps^3 \\  & 1 & \eps^2 \\  &   & 1\end{array}\right),
\eeq
where we have set $\eps=|V_{us}|=\sin\theta_C=0.226$ as the magnitude of the flavor spurion. This choice is taken to limit the tuning of $\OO(1)$ coefficients needed to explain hierarchies; if we had taken a smaller $\eps$, we would not have been able to explain the magnitude of $V_{us}$ without invoking a tuning of the parameters. If the horizontal symmetry is a direct product of multiple \uone's the magnitude of the spurions can be smaller, while still explaining hierarchies of order $V_{us}$ \cite{Leurer:1992wg,Nir:1993mx}; in section \ref{align}, we will be working in such a scenario.

To compute the mass hierarchies, we should take the magnitude of the Yukawa's at the flavor breaking scale $M$, and run them down to the observed, low-energy values; because the running does not alter significantly the mass ratios, we take the values of the running quark masses at $M_Z$, as listed in table \ref{masstable}; 
\begin{table}[tdp]
\begin{center}
\begin{tabular}{|c|c|c|c|c|c|c|c|c|}\hline
\ $u$ (MeV)&\ $d$ (MeV)&\ $c$ (MeV)&\ $s$ (MeV)& $b $ (GeV)& $t$ (GeV)&\ $e$ (MeV)&$\mu$ (MeV)&$\tau$ (MeV)
\\\hline
$1.3\pm.5 $&$ 2.9\pm1.2$&$620\pm80$&$\,55\pm15\,$&$\,2.89\pm0.09\, $&$\,171.7\pm3\,$&$0.487$&$102.7$&$1746$\\\hline
\end{tabular}
\end{center}
\caption{Running quark and charged lepton masses at $M_Z$, from ref. \cite{Xing:2007fb}. We have reported error bars only when they are sizable.}
\label{masstable}
\end{table}
%
%
\beq\label{hier}
m_t/v=1\sim\eps^0\,,\quad m_c/m_t\sim 0.0035\sim\eps^{4}\,,\quad m_u/m_c=0.002\sim\eps^{4},\qquad\ \ \  \\\nn
m_b/m_t=0.017\sim\eps^{2.7}\,,\quad m_s/m_b=0.019\sim\eps^{3}\,,\quad m_d/m_s=0.053\sim\eps^2,\quad \\\nn
m_\tau/m_t=0.01\sim\eps^{3.1} ,\quad m_\mu/m_\tau=0.059\sim\eps^{2}, \,\quad m_e/m_\mu=0.0047\sim\eps^{4}.
\eeq
Except for $b$ and $\tau$, for which the effect of $\tan\beta$ will be discussed below, rational numbers were approximated  by the closest integers. For example, the approximations $m_c/m_t=\eps^{3.8}\sim\eps^4$, or $m_s/m_b=\eps^{2.6}\sim\eps^3$ were taken. This is because the powers of $\eps$ for allowed Yukawa operators have to be integers and  an $\OO(1)$ coupling in front of any superpotential term will also contribute to the mass ratios; approximating $\eps^{0.5}\sim \eps^0$ is reasonable, because a factor of $\eps^{0.5}$ corresponds to have an $\OO(1)$ coefficient of $0.48$ in front of the relevant operator. It is also reasonable to approximate $m_s/m_b\sim\eps^2$; this only means that some of the charges might vary by $\pm1$, and it does not significantly change our conclusions.

For the down sector and the leptons, $\tan\beta$ has to be considered:
\beq
&&\frac{\vev{\phi_u}}{\vev{\phi_d}}=\frac{v_u}{v_d}=\tan\beta,\qquad v_u^2+v_d^2=v^2=(246 \text{ Gev})^2, \qquad \frac{m_b}{v_u}=\frac{m_b}{v_d}\frac1{\tan\beta},\nn\\
&&\implies\frac{m_b}{v_d}=\frac{m_b}{v_u}\tan\beta=\eps^{2-x_\beta},\qquad \frac{m_\tau}{v_d}=\frac{m_\tau}{v_u}\tan\beta=\eps^{2-x_\beta},
\eeq
where we defined $x_\beta=-0.7-\log_{\eps}(\tan\beta) >0$ to be an integer; values of $\tan\beta$ for which $x_\beta$ would not be exactly an integer can still be accounted for by considering the $\OO(1)$ couplings. For $\tan\beta\sim60$, we have $x_\beta=2$ and there is no hierarchy between the up and down sector. For small values, $\tan\beta\sim3$ corresponds to $x_\beta=0$, and $\tan\beta\sim1$  to $x_\beta=-1$.

\paragraph{Lepton sector:} For the leptons, the masses and mixing angles can be treated in a similar way:
\beq\label{lept}
Y^\ell_{ij}\sim v_d \eps^{\phi_d+L_i+\ell_j-r};\quad \frac{m^\ell_i}{m^\ell_j}\sim\eps^{L_i+\ell_i-L_j-\ell_j};\quad |U_{ij}|\sim\eps^{|L_i-L_j|}
\eeq
Here, $U$ is the PMNS mixing matrix; in principle, its expression in terms of $\eps$ depends on the specific mechanism chosen to generate neutrino masses. We assumed this form as it follows assuming either Dirac neutrino masses or a type I seesaw mechanism, as shown in the appendix. For $U$, almost all the elements are of order 1:
\beq
|U|=\left(
\begin{array}{ccc}
 0.82 & 0.55 & 0.16 \\
 0.36 & 0.7 & 0.62 \\
 0.44 & 0.46 & 0.77 \\
\end{array}
\right)
\sim\left(\begin{array}{ccc}1 & 1 & 1 \\1 &  1& 1 \\1 & 1 & 1\end{array}\right)
\quad\text{or}\quad \left(\begin{array}{ccc}1 & \eps & \eps \\ \eps & 1 & 1 \\ \eps &1  & 1\end{array}\right)
\quad\text{or}\quad \left(\begin{array}{ccc}1 & 1 & \eps \\ 1 & 1 & \eps \\ \eps &\eps  & 1\end{array}\right)
\label{pmns}.
\eeq
The choice of the PMNS parametrization changes the constraints that will be enforced on the charges of the leptons $L_i$; in  particular, the anarchical case corresponds to $L_{ij}=0$, while $|L_{12}|=|L_{13}|=1$ in the second case in eq. \eqref{pmns}. In the following, we will assume the anarchical mixing scenario, keeping in mind that there might be a difference of $\pm1$ in the $L_i$ charges if another hierarchy were to be generated by the horizontal symmetry. Although $|U_{13}|$ is small, it is still considered an $\OO(1)$ factor.

Putting together all the constraints \eqref{ckm}--\eqref{pmns}, the following relations must hold:
\beq\label{constraints}
\system {|Q_{12}|=1\\
|Q_{23}|=2\\
|Q_{13}|=3
}\qquad\qquad
\system{\phi_u+Q_3+u_3=r\\
Q_{23}+u_{23}=4\\
Q_{12}+u_{12}=4
}\qquad
\system{  \phi_d+Q_3+d_3=2-x_\beta+r\\
Q_{23}+d_{23}=3\\
Q_{12}+d_{12}=2
}\nn
\\
\system{
|L_{12}|=0\\
|L_{23}|=0\\
|L_{13}|=0
}\qquad\qquad
\system{
\phi_d+L_3+\ell_3=2-x_\beta+r\\
L_{23}+\ell_{23}=2\\
L_{12}+\ell_{12}=4
}\qquad\qquad\qquad\qquad
\eeq

These  are solved by two sets of solutions for the charge differences $\Phi_{ij}$:
\beq\label{diff}
\begin{array}{|ccc|ccc|ccc|ccc|ccc|}\hline
Q_{12}&Q_{13}&Q_{23}&d_{12}&d_{13}&d_{23}&u_{12}&u_{13}&u_{23}&L_{12}&L_{13}&L_{23}&\ell_{12}&\ell_{13}&\ell_{23}\\\hline
1&3&2&  1&2&1&  3&5&2&  \multirow{2}{*}{0}&\multirow{2}{*}{0}&\multirow{2}{*}{0}& \multirow{2}{*}{ 4} &\multirow{2}{*}{6} &\multirow{2}{*}{2}\\\cline{0-8}
-1&-3&-2&  3&8&5 &  5&12&7 &&&&&&\\\hline
\end{array}
\eeq
In the main body of the paper we will use the first set of solutions,
while the second one will be considered  in the appendix. The phenomenological implications are similar in both cases.
The remaining constraints are
\beq\label{3constr}
\phi_u+Q_3+u_3=r, \quad \phi_d+Q_3+d_3=2-x_\beta+r ,\quad \phi_d+L_3+\ell_3=2-x_\beta+r.\quad
\eeq
Because there are $17$ charges and only $13$ independent equations, they cannot be uniquely solved and the solutions depend on the choices of 4 independent variables, which we take as $\{Q_3, u_3, d_3, L_3\}$.

The $\mu$ term has a charge 
\beq
n_\mu=\phi_u+\phi_d-r=2-x_\beta-2Q_3-u_3-d_3+r.
\eeq
For positive charges of the fields, a $\mu$ term in the superpotential is easily avoided, as the right hand side of the equation becomes negative. In this case, the \Kahler-generated $\mu$ term is automatically suppressed with respect to the SUSY breaking scale $m_{3/2}$. For low-energy SUSY, $n_{\mu}$ should be a small negative number so that $\mu$ and $m_{3/2}$ are not that different. As stressed in the introduction, the anomaly cancellation conditions predict $\eps^{n_\mu}\sim\eps^{-1}$ and predicting $\mu=\eps m_{3/2}$. As we do not use a standard GS mechanism, $\nmu$ is not fixed in our framework; in particular, a rather heavy scale of SUSY breaking can still generate a weak scale $\mu$ term.

\subsection*{Additional freedom}
Until now, we have asked for the superpotential in eq. \eqref{supoY} to be invariant under the horizontal symmetry $U(1)_\HH$; but there are additional symmetries under which the superpotential is already invariant:
\beq
U(1)_B \times U(1)_L \times U(1)_Y \times U(1)_X.
\eeq
Those are the baryon number, the lepton number, the hypercharge and the Peccei-Quinn symmetry $U(1)_X$ under which $\phi_d$ has charge $-1$ and the $\bar d_i$'s and $\bar \ell_i$'s have charge $+1$. As pointed out in \cite{BenHamo:1994bq}, it is always possible to define a horizontal symmetry $\HH'$ related to $\HH$ by
\beq
\HH'=\HH+bB+lL+yY+xX,
\eeq
that still reproduces the hierarchies in eq. \eqref{hier}.
This results in different charges for the quarks and leptons that give the same charge differences \eqref{diff}; it should be noted that the four independent charges $Q_3, u_3, d_3, L_3$ can be shifted independently from each other with the action of the four additional symmetries.

\section{$R$-parity breaking}
\label{rpv}

In addition to the mass terms in the \supo, in the MSSM other dimension four operators are allowed that break the $R$-parity:
\beq\label{rps}
W_{\slashed R_p}=\bar \mu_i L_i\phi_u+\lambda_{ijk} L_iL_j\bar\ell_k +\lambda'_{ijk}L_iQ_j\bar d_k+\lambda_{ijk}''\bar u_i\bar d_j\bar d_k\,.
\eeq

The first three operators violate lepton number and the last violates baryon number. To have proton decay, both 
baryon number and lepton number violation are needed, while neutron and dinucleon decays are affected by $\lambda''$ on its own.
At the LHC, these operators allow sparticles to decay to SM fermions, generating events without missing energy. Then, most of the LHC limits on superpartners can be evaded and lighter squarks are allowed. 
Searches for $L$-violating decays of squarks and neutralinos are easier because final states include hard leptons. Searches for the $B$ violation involve a $\tilde t$ decaying to two jets through the coupling $\lambda''_{323}$, or a gluino decaying to three jets through a (possibly off-shell) stop.
Depending on the magnitude of the couplings $\lambda', \ \lambda''$, the LSP might be long-lived and a displaced vertex might be possible. The LSP lifetime might even be long enough for it to form a bound state and stop in the detector before decaying. We will consider the phenomenology at the end of this section.

In the presence of a \hors, the values of the RPV couplings  are determined by the horizontal charges of the superfields:
\beq
(\bar\mu_i, \lambda_{ijk},\lambda'_{ijk},\lambda''_{ijk})&\sim&\eps^{-r}(m\eps^{L_i+\phi_u}, \eps^{L_i+L_j+\ell_k},\eps^{L_i+Q_j+d_k}, \eps^{u_i+d_j+d_k}  ).
\eeq
Following an argument for $\bar\mu_i$ similar to the one about the $\mu$ term, $m$ is a scale that can be either $M$ or $m_{3/2}$ depending if the corresponding power of $\eps$ is positive or negative.

As we have seen above, requiring a \hors\,  does not mandate the  values of all the charges and it seems that it has no predictive power  for this sector; but, as the charge differences are determined in eq. \eqref{diff}, we can factorize the dependence on the unknown third generation charges and study the {\it relative} structure of the \rps \, coefficients\footnote{This property was first noted in \cite{Joshipura:2000sn}, and, in the presence of a GS anomaly cancellation, it was used to severely constrain the LNV operators.} (using the first solution for $Q_{ij}$ in eq. \eqref{diff}; for the second solution, see the appendix; also recall the notation $\Phi_{ij}=\Phi_i-\Phi_j$).
 The results for the different couplings are:
\begin{itemize}
\item bilinear LNV coupling $\bar\mu_iL_i\phi_u$:
\beq\label{2lnv}
\frac{\bar\mu_i}{\bar\mu_3}=\eps^{L_{i3}}=1,\qquad \bar\mu_1=\bar\mu_2=\bar\mu_3=m\eps^{n_{\bar\mu}}, \qquad \nbmu=L_3+\phi_u-r;
\eeq

\item trilinear LNV couplings $\lambda_{ijk}L_iL_j\bar\ell_k$ and  $\lambda'_{ijk}L_iQ_j\bar d_k$: first,
\beq\nn
\HH[\lambda_{233}]=-r+L_2+L_3+\ell_3=-r+L_2+Q_3+d_3=\HH[\lambda'_{333}] \implies \lambda_{233}\sim\lambda'_{333};\qquad
\eeq
The leading coefficients $\lambda_{233}$ and $\lambda_{333}'$ are the same (apart from $\OO(1)$ factors) and as such they are allowed or forbidden together. Defining $n_{LNV}=L_2+Q_3+d_3-r$, we have $\lambda_{233}=\lambda'_{333}=\eps^{n_{LNV}}$, and the textures of the coefficients are
\beq\label{3lnv}
\frac{\lambda_{ijk}}{\lambda_{233}}&=&\eps^{L_{i2}+L_{j3}+\ell_{k3}}=\eps^{\ell_{k3}},\qquad 
\left(\begin{array}{ccc} \lambda_{121}  & \lambda_{122}  & \lambda_{123}  \\ \lambda_{131} & \lambda_{132}  &\lambda_{133}   \\\lambda_{231}  & \lambda_{232}  & \lambda_{233} \end{array}\right)=\eps^{n_{LNV}}\left(\begin{array}{ccc}\eps^6 & \eps^2 & 1 \\ \eps^6 & \eps^2 & 1 \\ \eps^6 & \eps^2 & 1\end{array}\right),
\\
\frac{\lambda'_{ijk}}{\lambda'_{333}}&=&\eps^{L_{i3}+Q_{j3}+d_{k3}}=\eps^{Q_{j3}+d_{k3}},\qquad
\left(\begin{array}{ccc} \lambda'_{i11}  & \lambda'_{i12}  & \lambda'_{i13}  \\ \lambda'_{i21} & \lambda'_{i22}  &\lambda'_{i23}   \\\lambda'_{i31}  & \lambda'_{i32}  & \lambda'_{i33} \end{array}\right)=\eps^{n_{LNV}}
\left(\begin{array}{ccc} \eps^5&\eps^4&\eps^3\\\eps^4&\eps^3&\eps^2\\\eps&1&1\end{array}\right).\nn
\eeq

\item trilinear BNV coupling $\lambda''_{ijk}\bar u_i\bar  d_j \bar d_k$:
\beq\label{udd}
\frac{\lambda''_{ijk}}{\lambda''_{323}}=\eps^{u_{i3}+d_{j2}+d_{k3}},\qquad
\left(\begin{array}{ccc} \lambda''_{112}  & \lambda''_{212}  & \lambda''_{312}  \\ \lambda''_{113} & \lambda''_{213}  &\lambda''_{313}   \\\lambda''_{123}  & \lambda''_{223}  & \lambda''_{323} \end{array}\right)
=
\eps^{n_{BNV}} \left(\begin{array}{ccc}  \eps^7 &  \eps^4 & \eps^2  \\ \eps^6  &\eps^3   &   \eps \\  \eps^5&\eps^2   &1  \end{array}\right),
\eeq
where we have defined $\lambda''_{323}=\eps^{n_{BNV}}$, $n_{BNV}=u_3+d_2+d_3-r$.
\end{itemize}
It is worth noting that these textures are a general feature of any abelian horizontal symmetry as they come directly from the mass and mixing hierarchies and do not depend on other constraints such as anomaly cancellation.

It can be seen that the trilinear soft terms are of order $m_{3/2}$. For example, the BNV $A$-term comes from a superpotential term
\beq
\lambda''_{ijk}A_{\lambda''}=\int d^2\theta \frac X{M_P}\eps^{u_i+d_j+d_k} \bar u_i\bar d_j\bar d_k=m_{3/2}\eps^{u_i+d_j+d_k}\tilde{\bar u}_i\tilde{\bar d}_j\tilde{\bar d}_k=m_{3/2}\lambda''_{ijk} \tilde{\bar u}_i\tilde{\bar d}_j\tilde{\bar d}_k
\eeq
Thus, $A_{\lambda''}=m_{3/2}$. A similar computation applies to the other $A$-terms.

The coefficients $\bar\mu_3,\lambda_{233},\lambda'_{333},\lambda''_{323}$ are determined by the choice of the charges $Q_3,d_3,u_3,\\ L_3$. Because the individual charges vary under the additional action of a $U(1)_B \times U(1)_L \times U(1)_Y \times U(1)_X$ transformation, so will the overall coefficients; in particular, because the operators break $B$, $L$ and $X$, under a transformation 
\beq
bB+lL+xX
\eeq
the coefficients transform as $(\bar\mu,\lambda,\lambda',\lambda'')\to(\eps^l\bar\mu,\eps^{l+x}\lambda,\eps^{l+x}\lambda',\eps^{-b+2x}\lambda'')$,  or
$$
(n_{\bar\mu},n_{LNV},n_{BNV})\to(n_{\bar\mu}+l,n_{LNV}+l+x,n_{BNV}-3b+2x).
$$
This means that they are not fixed by the fermion hierarchies, as their values can be shifted by a $B$, $L$ or $X$ transformation. Hence, as first pointed out in ref. \cite{BenHamo:1994bq}, if $l,b,x$ are large enough, the \rps\ couplings will just be too small to be of any phenomenological significance; if one considers that arbitrarly high values for the charges are unnatural and limits them to be at most of order 10, proton decay constraints can still be satisfied \cite{BenHamo:1994bq}. In the next section we will see that, while this argument still holds for heavy superpartner masses, it does not for sub-TeV SUSY: {\it arbitrarly small \rps\ coefficients would either mimic $R$-parity conserving supersymmetry, or allow the formation of $R$-hadrons or other stable massive particles.}

We can trade the 4 independent charges $(Q_3,u_3,d_3,L_3)$ with the `phenomenological' variables $(Q_3,n_\mu,\nbmu,n_{BNV})$ (in the sense that they determine the \rps\, phenomenology of the models): using the constraints \eqref{3constr}, it can be seen that $n_{LNV}$ is related to these variables by
\beq
\nmu=\nbmu-n_{LNV}+2-x_\beta+r.
\eeq
For a weak scale $\mu$ term, $\nmu$ will need to be a negative integer number of order 1, so that $\mu=\eps^{|\nmu|}m_{3/2}$. $\nmu$ cannot be fractional, or the $\mu$ term would not be generated at all. Then, the coefficients $\nbmu$ and $n_{LNV}$ are either both integer or both fractional.

In terms of the variables $(\nbmu,n_{BNV})$, we have several phenomenological scenarios:\footnote{A similar classification allowing/forbidding leptonic or baryonic RPV was outlined in refs. \cite{Sierra:2009zq,Florez:2013mxa}.}
\begin{enumerate}
\item If they are both fractional, all the dimension four RPV operators are forbidden;  LHC searches relying on missing energy apply and the weak scale is generically fine-tuned.

\item If they are both integers, neither $B$ or $L$ are conserved; the proton can decay to mesons and leptons, through a product of $\lambda'$ and $\lambda''$ couplings. The leading constraints from upper limits on nucleon lifetimes are \cite{Barbier:2004ez}:
\beq
p\to \pi^0\ell^+:&\ &|\lambda'_{l1k}{\lambda''}^*_{11k}|\lesssim 2\times 10^{-27} \left(\frac{m_{\tilde d_{kR}}}{100\text{ GeV}}\right)^2\nn\\
p\to K^+\bar\nu:&\ &|\lambda'_{i2k}{\lambda''}^*_{11k}|\lesssim 3\times 10^{-27} \left(\frac{m_{\tilde d_{kR}}}{100\text{ GeV}}\right)^2\\
n\to \pi^0\bar\nu: \ &\ &|\lambda'_{31k}{\lambda''}^*_{11k}|\lesssim 7\times 10^{-27} \left(\frac{m_{\tilde d_{kR}}}{100\text{ GeV}}\right)^2\nn
\eeq
Substituting the expressions for the couplings in terms of $\eps$, the leading constraint is the second one:
\beq\label{proton}
|\lambda'_{i23}{\lambda''}^*_{113}|=\eps^{n_{LNV}+n_{BNV}+8}\lesssim10^{-{27}} (m_{\tilde b_{R}}/100\text{ GeV})^2= \eps^{41}( m_{\tilde b_{R}}/100\text{ GeV})^2.\qquad
\eeq
A priori, this is possible if both $n_{BNV}$ and $n_{LNV}$ are of order 17 or higher, forcing the charges of the individual fields to be of order 10, as considered in \cite{BenHamo:1994bq}. As we will discuss in section \ref{pheno}, the individual couplings are very small  and either give missing energy events at the LHC or heavy particles that are stable on collider timescales. In both cases, generic limits for the sparticles masses go up to and above 1 TeV  and this scenario can be neglected when considering low-energy SUSY.

\item If only $n_{BNV}$ is fractional there is no $\slashed B$ operator, while lepton number violation is allowed for the three operators $L\phi_u,\ LL\bar\ell,\ LQ\bar d$ (an interesting case opens for decaying dark matter neutralinos, studied in ref. \cite{Sierra:2009zq}). These interactions usually give rise to collider signatures including multiple leptons and searches at the LHC exist for a stop LSP \cite{CMS-PAS-SUS-13-003}. In this scenario, the limits are near or above a TeV and would rule out a low SUSY scale for a considerable portion of the parameter space.

\item If only $\nbmu$ is fractional, so is $n_{LNV}$ and there is no lepton number violation. The only RPV operator is $\bar u\bar d\bar d$, which allows a stop LSP to decay to jets and no missing energy. The idea of a  $\bar u\bar d\bar d$ operator has seen a revival since the null LHC searches and its phenomenology has been studied \cite{Allanach:2012vj}; it arises in several models of low-energy SUSY that evade the LHC bounds \cite{Csaki:2011ge,Bhattacherjee:2013gr,Florez:2013mxa}. We will consider this scenario and put bounds on the magnitude of $\lambda''$ in section \ref{pheno}.
\end{enumerate}

\subsection{dimension five operators}\label{d5}
Although $R$-parity is usually assumed to make the proton stable by forbidding the dimension four operators in \eqref{rps}, there are dimension five operators that have to be suppressed to avoid  proton decay: they come from both superpotential and \Kahler corrections \cite{Dreiner:2012ae}:
\beq\label{d5op}
W_5&=&\frac{(\kappa_1)_{ijkl}}{M_P}{Q}_i{Q}_j{Q}_k{L}_l+\frac{(\kappa_2)_{ijkl}}{M_P}{\bar u}_i{\bar u}_j{\bar d}_k{\bar \ell}_l+\frac{(\kappa_3)_{ijk}}{M_P}{Q}_i{Q}_j{Q}_k{\phi}_d+\frac{(\kappa_4)_{ijk}}{M_P}{Q}_i{\phi}_d{\bar u}_j{\bar \ell}_k+\nn
\\&&+\frac{(\kappa_5)_{ij}}{M_P}{L}_i{\phi}_u{L}_j{\phi}_u+\frac{(\kappa_6)_{i}}{M_P}{L}_i{\phi}_u{\phi}_d{\phi}_u,\\
K_5&=&\frac{(\kappa_7)_{ijk}}{M_P}{\bar u}_i{\bar d}^{*}_j{\bar\ell}_k+\frac{(\kappa_8)_{i}}{M_P}{\phi}^{*}_u{\phi}_d{\bar \ell}_i+\frac{(\kappa_9)_{ijk}}{M_P}{Q}_i{L}^{*}_j{\bar u}_k+\frac{(\kappa_{10})_{ijk}}{M_P}{Q}_i{Q}_j{\bar d}^{*}_k\nn
\eeq
Some of these operators break $B$, some break $L$, and some break both; 
generically, these operators are dangerous for nucleon decay because they generate an effective operator $\eta_{eff}qqq\ell$ (in the same way as $LQ\bar d$ and $\bar u\bar d\bar d$ do if they are both allowed) where proton stability requires $\eta_{eff}\lesssim10^{-32} \text{GeV}^{-2}$.
For example, for the operator $\frac1{M_P}QQQL$, $\eta_{eff}\sim \frac1{M_P M_{SUSY}}\sim 10^{-21}\text{ GeV}^{-2}$, and an additional suppression of order $10^{-10}$ is needed. 
In the $R$-parity conserving MSSM, there is usually no explanation for this small factor. In a grand unified theory where the effective operator arises from a colored Higgs exchange, $\eta_{eff}$ is proportional to the Yukawa couplings of the lighter generations, providing enough suppression. In a \rps \ model with horizontal symmetries, the couplings can be suppressed in the same way all the other couplings are, by different powers of $\eps$.

\begin{table}[tdp]
\begin{center}
\[
\begin{array}{|c|c|c|c|} \hline
\OO_1 &6+r-2x_\beta-n_{BNV}-2\nmu+\nbmu &
\OO_2 &-2-r+x_\beta+n_{BNV}+\nmu-\nbmu \\\hline
\OO_3 &6+r-2x_\beta-n_{BNV}-\nmu &
\OO_4 & 2-x_\beta+r-\nbmu \\\hline
\OO_5 & 2\nbmu +2r &
\OO_6 &\nmu+\nbmu+2r \\\hline
\OO_7 & -\nbmu &
\OO_8 & 2-x_\beta-\nbmu \\\hline
\OO_9 & -\nbmu&\OO_{10} &
 2-x_\beta-n_{BNV}-\nmu \\\hline
\end{array}
\]
\end{center}
\label{d5table}
\caption{Horizontal charges $\HH[\OO_i]$ of the leading component of the operator $\kappa_i \OO_i$ in eq. \eqref{d5op}. The leading component is the one with the greatest number of third generation fields allowed by the antisymmetric combinations of fields mandated by the gauge structure: for example, for $\OO_1$ the leading component is $(Q_2Q_3)(Q_3L_3)$. The numerical value of the coefficient $\kappa_i$ is $\eps^{\HH[\OO_i]-r}$.}
\end{table}

As in the previous section, we can factorize the leading coefficient for each operator and use the solutions for the SM charge differences \eqref{diff} to find textures for each operator. We will not write the textures down as they  follow from the same considerations as above and are of limited phenomenological interest; our main interest will be to find the leading coefficient and to see if the whole operator is allowed or forbidden: the charges of  (the leading component of) each operator are given in table \thetable. 

In particular, if $\nbmu$ is fractional all the $\Delta L=1$ operators ($\OO_1,\OO_2,\OO_4,\OO_6,\OO_7,\OO_8,\OO_9$) are forbidden (a similar relationship between some dimension five operators was noted in a different context in \cite{Dreiner:2012ae}). $\OO_5$ (Weinberg's neutrino mass effective operator) is allowed if $\nbmu$ is a half integer. This gives an irreducible contribution to the neutrino masses, as such an operator has to be generated at $M_P$ (even if we do not have a specific neutrino sector in mind). Of course, it is possible that the effective operator is also suppressed by a smaller scale (as in the seesaw mechanism), thus generating sizeable neutrino masses.
 The other operators left for a fractional $\nbmu$ are $\OO_3$ and $\OO_{10}$, who do not need additional suppression as they could mediate proton decay only if combined with  dimension four $\slashed L$ operators, which are forbidden for fractional $\nbmu$.

For operators of dimension five or higher,  negative powers of $\eps$ can arise: imagine that a heavy field $\Phi$ (charged under the SM group) acquires a mass proportional to $S$, $m\propto {S}$, and  that it has trilinear couplings to the light fields. Then, one can get dimension five operators for the light fields from a superspace diagram where the internal propagator is the heavy field; its propagator reads $\frac{m^*}{p^2-|m|^2}\sim\frac1S$ and  a negative power of $\eps$ is present in the low energy theory. This does not change our conclusions, as operators with fractional powers of $\eps$ are still forbidden, and no dimension five operator has been forbidden because it had a negative power of $\eps$.

\subsection{Phenomenology}\label{pheno}

In this section, we will study limits on the \rps \ couplings in the superpotential; assuming that the coefficient $\nbmu$ is a half integer, all the $L$-violating dimension four and 5 operators are forbidden, apart from the neutrino mass term. We are left with the $B$-violating superpotential
\beq
W_{\slashed B}=\lambda_{ijk}''\bar u_i\bar d_j\bar d_k.
\eeq

The main motivation for our choice is to avoid stringent limit like the proton decay bound \eqref{proton} while still considering low-energy supersymmetry. Then we cannot have arbitrary small coefficients in the \rps\  superpotential of eq. \eqref{rps}: if a  LSP is produced at the LHC but cannot decay because its \rps \ coupling is too small, it will either exit the detector as missing energy if it is neutral (therefore, the limits on R-parity conserving SUSY will apply), or hadronize and be observed as a new stable massive particle (an R-hadron). R-hadrons have been investigated by ATLAS and CMS, and they exclude a stop LSP up to 680 GeV \cite{ATLAS-CONF-2012-075} and 850 GeV \cite{CMS:hwa}, respectively.

With these limits, we can exclude a range of \rps\  couplings that would make the LSP stable on collider timescales: the width and decay length of a stop decaying directly to two quarks through the coupling $\lambda''_{3ij}$ are:
\beq\label{stopdecay}
&&\Gamma(\tilde t\to d_id_j)=\frac{m_{\tilde t}}{8\pi}\sin^2\theta_{\tilde t}|\lambda_{3ij}''|^2, \\
&&c\tau=\frac1210^{-16}|\lambda''_{3ij}|^{-2}\left( \frac{100 \text{ GeV}}{m_{\tilde t}}\right) \text{m}\sim\eps^{27-2n_{BNV}}\left( \frac{850 \text{ GeV}}{m_{\tilde t}}\right) \text{m}\nn.
\eeq
where we have taken the coupling $\lambda''_{323}$ and assumed maximal mixing.\\
For $n_{BNV}\gtrsim13$, the decay length is bigger than $1$ m and a $\tilde t$ LSP would form an R-hadron which stops or decays within the detector. Allowing a {\it natural} stop with $m_{\tilde t} < 850$ GeV requires $n_{BNV}<13$ (or correspondingly, $\lambda''\gtrsim10^{-8}$).
A similar order of magnitude  bound applies to the LNV couplings $\lambda'$, if they were allowed. Consequently, the proton decay rate, proportional to $\lambda'_{i23}\lambda''_{113}$, would be $10^9$ times faster than the experimental limits.
This is why we do not consider the possibility of both BNV and LNV violation in the renormalizable superpotential.

In ref. \cite{Evans:2012bf}, stricter limits are inferred from more complicated decay topologies, e.g. a stop decaying to a top and a neutralino $\tilde X$, which gives another sfermion $\tilde f$ ultimately decaying through a RPV coupling $C$ to two fermions. The resulting decay length is
\beq
c\tau\sim2\times 10^{-10}|C|^{-2}\left(\frac{0.01}{\alpha}\right)^2\left(\frac{m_{\tilde X}}{m_{\tilde t}}\right)^2\left(\frac{m_{\tilde f}}{m_{\tilde{t}}}\right)^4\left( \frac{850 \text{ GeV}}{m_{\tilde t}}\right)\text{m}
\eeq
To have $c\tau\lesssim 1$m, we need $C$ to be of order $10^{-5}$ or greater. If $\tilde f$ is a sbottom and $C$ is $\lambda''_{ij3}$, this corresponds to a stricter bound, $n_{BNV}\lesssim 6$. This is limit is not relevant in our case because the amplitude for the direct decay \eqref{stopdecay} would always be bigger and the stop would dominantly decay through that channel. In ref. \cite{Evans:2012bf}, only one coupling at a time was assumed to be dominant, which in our case would correspond to having $n_{BNV}<0, \ \lambda''_{323}=0$ and some other $\lambda''_{ijk}\neq0$. As we will see later, for our scenario this is not compatible with the bounds from low-energy physics.

ATLAS and CMS can reconstruct displaced vertices for hadronic stops decaying 1mm $-$ 10cm away from the interaction point (for a similar discussion in a different $R$-parity violating  model, see \cite{Krnjaic:2013eta}). A stop would generate a displaced vertex for the range $11\lesssim n_{BNV}\lesssim 13$. Ruling out these displaced decays would restrict the remaining parameter space of natural supersymmetric models with horizontal symmetries and \rpv.

We can now review how low-energy decays put an upper limit on the couplings $\lambda''$, and therefore a lower bound on the coefficient $n_{BNV}$; most of the expressions for the limits come from the review \cite{Barbier:2004ez} on $R$-parity violating SUSY. These bounds depend generically on squarks and gluino masses, which we will be assuming to be at a common scale (although not degenerate) $\tilde m\sim m_{\tilde q_k}\sim m_{\tilde g}$. A factor of a few in these relations would not change the results significantly.

Comparable bounds come from neutron-antineutron oscillation and dinucleon decay, while  limits from $B$ physics are subdominant: here, we update the neutron-antineutron oscillation period $\tau_{n-\bar n}$ of ref. \cite{Barbier:2004ez} with the latest lower limit by the SuperKamiokande experiment \cite{Abe:2011ky}, $\tau_{n-\bar n}>2.44\times 10^8$s.
\begin{itemize}

\item  dinucleon decay $NN\to KK$: from \cite{Litos:2010zra,Csaki:2011ge,Florez:2013mxa} we read the limit
\beq
|\lambda''_{112}|\lesssim3\times 10^{-7}\left(\frac{1.7\times10^{32}\, \text{yr}}{\tau_{NN\to KK}}\right)^{1/4}
\left(\frac{m_{\tilde s_{R}}}{300\, \text{GeV}}\right)^{2}\left(\frac{m_{\tilde g}}{300\, \text{GeV}}\right)^{1/2}
\left(\frac{75\text{ MeV}}{\tilde\Lambda}\right)^{5/2}
\eeq
where $\tilde \Lambda$ is the hadronic scale arising from the hadronic matrix element and phase-space integrals. Thus, we have
\beq
 |\lambda''_{112}|=\eps^{n_{BNV}+7}\lesssim \eps^9\left(\frac{\tilde m}{500GeV}\right)^{5/2} ,\qquad n_{BNV}\gtrsim 2.
\eeq
As argued in \cite{Litos:2010zra}, given the weak dependence of the limit from $\tau_{NN\to KK}$, this channel's limit will likely not increase substantially in the future.

\item $n-\bar n$ oscillation: the limit cited in \cite{Barbier:2004ez} is
\beq\label{nn}
|\lambda''_{11k}|\lesssim (10^{-8}-10^{-7})\frac{10^8s}{\tau_{n-\bar n}}\left(\frac{m_{\tilde q_{kR}}}{100\text{ GeV}}\right)^{2}\left(\frac{m_{\tilde g}}{100\text{ GeV}}\right)^{1/2},
\eeq
which in our model of horizontal symmetries, reads (for $k=3$)
\beq
|\lambda_{113}''|=\eps^{n_{BNV}+6}\lesssim\eps^{10}\left(\frac{\tilde m}{500\text{ GeV}}\right)^{5/2} ,\qquad n_{BNV}\gtrsim4.
\eeq
For this process, the original calculation by Zwirner \cite{Zwirner:1984is} assumed an unknown $10\%$ LR mixing in the squark mass matrix: for a non-$R$ horizontal symmetry, the soft terms give  $\tilde M^2_{LR}/\tilde M_{LL}^2\sim M/\tilde m$, where $M$ is the squark mass matrix \cite{Leurer:1993gy}. In a $n-\bar n$ oscillation involving a sbottom, the contribution is of order $m_b/\tilde m\sim 1/100$ for weak scale supersymmetry (and $10^{-3}$ for TeV scale SUSY), which is a factor of 10 (respectively, 100) smaller than the value assumed by \cite{Zwirner:1984is}. The bound changes accordingly: for $\tilde m\lesssim 1$ TeV we get $n_{BNV}\gtrsim2$ instead of $n_{BNV}\gtrsim4$. For a horizontal $R$-symmetry, the left-right mixing is large \cite{Leurer:1993gy} and the stricter bound applies.

Alternative mechanisms for neutron oscillation that do not involve unknown soft terms have been proposed in \cite{Goity:1994dq,Chang:1996sw}, and the bounds are similar:
\beq
|\lambda_{313}|=\eps^{n_{BNV}+1}\leq 10^{-2} =\eps^3 ,\qquad  n_{BNV}\gtrsim 2
\eeq

\item neutron decay $n\to \Xi$: 
\beq
|\lambda''_{112}|&\lesssim& 10^{-8.5}\left(\frac{m_{\tilde g}}{100\text{ GeV}}\right)^{1/2}\left( \frac{m_{\tilde s_{R}}}{100\text{ GeV}}\right)^2 \left(\frac{10^{32}yr}{\tau_{NN}}\right)^{1/4}\left(\frac{10^{-6}\text{ GeV}^6}{\langle{\bar N}|{ududss}|{\Xi}\rangle}\right)^{1/2}\nn
\\
|\lambda''_{112}|&=&\eps^{n_{BNV}+7}\lesssim\eps^{10}\left(\frac{\tilde m}{500\text{ GeV}}\right)^{5/2},\qquad n_{BNV}\gtrsim3.
\eeq
\item $B^-\to \phi\pi^-$ decay:
\beq\label{bb}
|{\lambda''}^*_{i23}\lambda''_{i12}|\lesssim 6\times 10^{-5} \left(\frac{m_{\tilde u_{iR}}}{100\text{ GeV}}\right)^2
,\qquad 
|{\lambda''}^*_{323}\lambda''_{312}|=\eps^{2n_{BNV}+1}\lesssim \eps^5 \left(\frac{m_{\tilde t_{R}}}{500 \text{ GeV}}\right)^2 \nn
\eeq
from which we conclude $n_{BNV}\gtrsim2$.
\end{itemize}
The stronger limit comes from the neutron decay channel  $n\to \Xi$, and it is $n_{BNV}\gtrsim3$.

The limits can be compared with the MFV prediction for the magnitude of the $\bar u\bar d \bar d$ couplings (MFV \rps\ SUSY was introduced in ref. \cite{Nikolidakis:2007fc} before the start of the LHC; for its implications on the LHC limits on the scale of supersymmetry, see ref. \cite{Csaki:2011ge}). In this scenario, the Yukawa matrices are considered as spurions of the flavor symmetry group $SU(3)^5$, and only terms that are invariant under this symmetry are allowed. Thus, the operator $\bar u\bar d \bar d$ is neutral under the flavor group only if accompanied by products of Yukawa matrices, and it reads $Y_uY_dY_d \bar u\bar d \bar d$. The couplings involving light quarks are suppressed with respect to the couplings involving the top. 

We can compare the MFV prediction to the horizontal symmetry results (as was first done in ref. \cite{Florez:2013mxa}), taking  the numerical results in ref. \cite{Csaki:2011ge} expressed  as powers of $\eps$:\footnote{The couplings depend on the SUSY breaking scale, at which the Yukawas (and quark masses) should be evaluated. As the running of the quark masses between $M_Z$ and $m_{3/2}$ does not  change the exponents, we evaluate them at $M_Z$. We thank the authors of ref. \cite{Florez:2013mxa} for helping us correct an earlier version of our computation.}
\beq
 \lambda''_{MFV}=\eps^{11+2\log_\eps \tan\beta}\left(\begin{array}{ccc}\eps^{13} & \eps^8 &\eps^ 2 \\\eps^8 & \eps^3 & \eps^1 \\\eps^5 & \eps^2 & 1\end{array}\right).
\eeq
Using our notation, the largest coupling is $\lambda_{323}''=\eps^{n_{BNV}}$ with $n_{BNV}=11+2\log_\eps \tan\beta$, while the structure of the matrix is slightly different compared to the horizontal symmetry prediction, eq. \eqref{udd}. The largest coupling is still $|\lambda''_{323}|$, but there is more suppression for the couplings involving lighter quarks. The collider phenomenology is similar \cite{Nikolidakis:2007fc,Csaki:2011ge,Florez:2013mxa} and for the most part corresponds to prompt decays. Displaced vertices are allowed only for $\tan\beta<10$, corresponding to $|\log_\eps \tan\beta|< 1.5$ and $n_{BNV}>8$; in our framework, there is no connection between $\tan\beta$ and the possibility of having displaced vertices.
Extreme values of $\tan\beta\sim100$ would bring the exponent of $\lambda''_{323}$ as low as  $n_{BNV}=5$.

\bigskip

To summarize this section, generic \rpv\ with both lepton number and baryon number violation is inconsistent with the absence of superpartners that are stable on collider scales. Low energy supersymmetry where baryonic \rpv \ is combined with a horizontal symmetry is only allowed in the range $3\lesssim n_{BNV}\lesssim 13$, where $\eps^{n_{BNV}}$ is the magnitude of the biggest $R$-parity violating operator, $\lambda''_{323}\bar u_3\bar d_2\bar u_3$. Currently, the most relevant LHC searches for this model are performed by ATLAS: 
\begin{itemize}
\item In \cite{ATLAS:2012dp} a pair produced massive particle decaying to three jets is looked for: this topology can describe a decay chain of a gluino to three quarks,  $\tilde g\to q\tilde q \to qqq$, where the last decay involves an RPV coupling. Assuming an off-shell squark, the $95\%$ confidence level limit on the gluino mass is $m_{\tilde g}>666 $ GeV.
\item In \cite{ATLAS:2013tma}  two same-sign leptons in the final state are searched for, as a signature of two gluinos decaying as $\tilde g\to \bar t\tilde t\to\bar t bs$. The 95\% confidence level limit on the gluino mass is $m_{\tilde g}>890 $ GeV. 
\end{itemize}
Unfortunately, these searches do not give stringent limits on the stop mass, which is most important when thinking about naturalness of the weak scale. Still, the gluino mass enters the RGE of the stop mass term, so that if in the near future a gluino  is excluded above $1.4$ TeV, the fine tuning of the weak scale would be less than 1\% \cite{Feng:2013pwa}.

\section{Quark-Squark Alignment}
\label{align}

In the last section we have seen how $R$-parity violation can solve the tension between the negative LHC searches for supersymmetry and the presence of light superpartners.

However, it has long been known  that a generic low-energy supersymmetric spectrum generates unobserved FCNCs. In particular, neutral meson oscillations are well explained by the Standard Model, leaving little space for  new physics contributions. The flavor structure of a supersymmetric extension of the SM has to be highly non-trivial. If squark degeneracy is assumed, FCNCs can be suppressed. However, this is a strong assumption, and does not follow automatically from abelian horizontal symmetries: we will here focus on aligned models \cite{Leurer:1993gy}, in which the quark and squark mass matrices are diagonal in the same basis in which the gluino interactions are diagonal. Similarly to squark degeneracy, alignment suppresses FCNCs in $K-\bar K$ and $B-\bar B$ oscillation. It turns out that the squark bases cannot be aligned for both the up and down sector and that, with TeV scale SUSY, an $\OO(10\%)$ squark degeneracy is still needed to explain the observed $D-\bar D$ mixing  \cite{Gedalia:2009kh}.

A natural way to get aligned models is to use horizontal symmetries \cite{Nir:1993mx,Leurer:1993gy}: in particular, a simple model involves two symmetries $\HH_1=U(1)_{\HH_1}$ and $\HH_2=U(1)_{\HH_2}$ with two spurions $\eps_1, \ \eps_2$ carrying charges $(-1,0)$ and $(0,-1)$ under $(\HH_1,\HH_2)$. The fermion mass hierarchies and mixings can be reproduced and at the same time the sfermion mass matrix can be non-generic, suppressing flavor changing neutral currents. 

In a model with two horizontal symmetries, we can see that the \rps\  couplings are either the same as calculated earlier in eqs. \eqref{2lnv}--\eqref{udd}, or zero; to see this, we define coefficients $a$ and $b$ such that $\eps_1=\eps^a, \eps_2=\eps^b$, with $ \eps=0.226$ (in this case, the spurions are smaller).  The theory is then invariant under the diagonal subgroup $\HH_d\subset \HH_1\times \HH_2$ with a spurion $\eps$ and charges $\HH_d[\Phi]=a\HH_1[\Phi]+b\HH_2[\Phi]$. A generic operator $\OO$ in the superpotential with charges $p,q$ under $\HH_1,\HH_2$ will be suppressed by
\beq
\eps_1^p\eps_2^q\OO=\eps^s\OO, \qquad s=ap+bq.
\eeq
The charge under $\HH_d$ is a linear combination of the charges under $\HH_1,\HH_2$. It is important to note that $s$ can be positive even though $p$ or $q$ might be negative; an operator that is forbidden by the full symmetry would be allowed by the subgroup $\HH_d$. In the same way, an operator that has an integer charge under $\HH_d$ might have a fractional charge under $\HH_1$ or $\HH_2$, and therefore be forbidden. For this reason, when there are two or more symmetries, the \rps\ couplings can either be zero or be determined by the textures in eqs. \eqref{2lnv}--\eqref{udd}. It is possible to write down aligned models for both cases.


For a specific example of how an \rps\ operator can be forbidden by $\HH_1\times \HH_2$ even if it is allowed by the diagonal $\HH_d$, we take an aligned model  from \cite{BenHamo:1994bq}, with $a=1, b=2$ and the following charges for the lepton sector:
\beq
\begin{array}{c|ccc|ccc}  & L_1& L_2  & L_3  & \ell_1  & \ell_2  &\ell_3     \\\hline \HH_1 & 5& -1& 1& -3& 2& 0  \\\hline  \HH_2 &  1& 4& 3& 5& 1& 1  \\\hline\hline
\HH_d&7&7&7&7&4&2
\end{array}
\eeq
For the \rps\ term $\lambda_{ijk}L_iL_j\ell_k$, one can factor out  $\lambda_{233}=\eps_2^8=\eps^{16}$, and compute the textures of $\lambda_{ijk}/\lambda_{233}$. If we just considered $\HH_d$ the texture would be
\beq
(\lambda_{ij1},\lambda_{ij2},\lambda_{ij3})=\eps^{16} (\eps^5,\eps^2,1).
\eeq
Instead, using the full symmetry $\HH_1\times \HH_2$, the coefficients are
\beq
(\lambda_{ij1},\lambda_{ij2},\lambda_{ij3})=\eps_2^{8}\ (\eps_1^{-3}\eps_2^4,\eps_1^2,1)= (0,\eps_2^{8}\eps_1^2,\eps_2^{8}).
\eeq
As $\lambda_{ij1}$ has a negative power of $\eps_1$, it is a holomorphic zero. A specific coupling can be forbidden, while the others maintain the previous structure. For a different choice of the horizontal charges one can  build a model where the whole superpotential term is forbidden due to negative charges, or a model in which some operators have fractional charges with respect to $\eps_1$ or $\eps_2$. 

This does not changes the phenomenological conclusions of section \ref{pheno}, as  one can get some of the couplings to be zero, but, in general, similar bounds will be generated from the remaining non-zero coefficients. We conclude that aligned models of \rps\ SUSY with  horizontal symmetries are subject to the same order of magnitude limits that we computed previously.


\section{Horizontal symmetry implications for the NMSSM}
\label{NMSSM}

In the presence of light stops, the Higgs mass needs an extra contribution to reach the measured value of 126 GeV. In the Next-to-Minimal Supersymmetric Standard Model (NMSSM)\cite{Ellwanger:2009dp}, an extra singlet $N$ has a tree-level coupling to the Higgs doublets. It should be noted that in our scenario the singlet is needed just to raise the Higgs mass while keeping light stops, and the arguments in the other sections are not influenced by choosing an alternative mechanism.\footnote{In the NMSSM, a $\mu$ term is automatically generated. As seen earlier, a $\mu$ term can also be generated by a \Kahler correction in a Giudice-Masiero mechanism: this does not happen in the $Z_3$ symmetric NMSSM, but it does in the general NMSSM. The interplay between these two mechanism is left for discussion in a future work.} We first review the model, and then see the horizontal symmetry constraints on the NMSSM.
We take the superpotential involving $N$ as:
\beq\label{nmssmsup}
W \supset \lambda N\phi_u\phi_d +\frac{\kappa}3N^3,
\eeq
where for simplicity we are taking the so called $Z_3$ symmetric NMSSM \cite{Ellis:1988er,Ellwanger:2009dp}, where all the dimensionful couplings have been put to zero. With this superpotential, the $\mu$ term is generated dynamically, $\mu=\lambda \vev N$; in addition, the Higgs mass receives an extra contribution, which in the limit $\kappa \vev N\gg |A_\kappa|,|A_\lambda|$, can be written as
\beq
 m_h^2&\simeq& M_Z^2\cos^2 2\beta+ \lambda^2 v^2\sin^2 2\beta -\frac{\lambda^2 v^2}{\kappa^2}(\lambda-\kappa\sin2\beta)^2\nn\\
 &+&\frac{3m_t^4}{4\pi^2v^2}\left(\log \frac{m_{\tilde t}^2}{m_t^2}+\frac{A_t^2}{m_{\tilde t}^2}\left(1-\frac{A_t^2}{12m_{\tilde t}^2}\right)\right)
\eeq
where the first line includes the additional tree-level contribution proportional to $\lambda$ and the second line is the usual stop loop contribution present in the MSSM. This expression is maximized for $\lambda\sim0.7$ and $\tan\beta\sim 2$. A Higgs mass of 126 GeV can be achieved with  stops around 500 GeV and $\tan\beta\sim 2$, and result in moderately low fine-tuning of the weak scale \cite{Hall:2011aa}.

The soft SUSY breaking terms are
\beq
V_{soft}\supset m_N^2|N|^2+\lambda A_\lambda N \phi_u\phi_d +\frac13 A_\kappa \kappa N^3
\eeq
and for large $N$, an absolute minimum of the superpotential is found for $A_\kappa^2>9m_N^2$ at 
\beq
\vev N=\frac1{4\kappa}\left(-A_\kappa-\sqrt{A_k^2-8m_N^2}\right)
\eeq

\bigskip

With a horizontal symmetry, the terms in the superpotential \eqref{nmssmsup} must be neutral under $\HH$: because we are taking $\lambda\sim0.7$ to maximize the Higgs mass, $\lambda$ is assumed to be an $\OO(1)$ factor, that is, carrying a null horizontal charge. This fixes the charge of $N$:
\beq
\HH[N]=-\phi_u-\phi_d+r=-\nmu.
\eeq
In particular, a negative $\nmu$ translates in a positive $\HH[N]$. We will denote $\HH[N]$ by $N$. The constant $\kappa$ is fixed as $\kappa\sim\eps^{3N-r}$. The soft terms are of order $m_{3/2}$, as they come from non-renormalizable corrections as $\int d^4\theta NN^\dag \left(1+\frac{XX^\dag}{M_P^2}\right)$, $\int d^2\theta \frac X{M_P} N\phi_u\phi_d$ and $\int d^2\theta \frac X{M_P} N^3$. The minimum for $N$ is unchanged.

Finally, there are some additional operator involving the singlet field $N$: the only renormalizable operator is
\beq
\bar\lambda_i NL_i\phi_u
\eeq
If allowed, this would generate an effective $\bar\mu_i$ term with magnitude $\bar\mu_i=\bar\lambda_i \vev N$, with $\bar\lambda_i=\eps^{N+L_i+\phi_u-r}=\eps^{\nbmu-\nmu}\eps^{L_{i3}}$. For the scenario with no lepton number violation, with $\nbmu$ semi-integer, the operator is forbidden.

There are also dimension 5 operators involving $N$:
\beq\label{d5Nop}
W_{5,N}&=&\frac{(\kappa_{11})_{i}}{M_P}L_i\phi_u NN+\frac{(\kappa_{12})_{ijk}}{M_P}L_iL_j\ell_k N
+\frac{(\kappa_{13})_{ijk}}{M_P} L_iQ_j\bar d_k\nn
+\frac{(\kappa_{14})_{ijk}}{M_P}\bar u_i\bar d_j\bar d_k N\\
K_{5,N}&=&\frac{(\kappa_{15})_{i}}{M_P} L_i\phi_uN^*
\eeq
\begin{table}[tbdp]
\begin{center}
\[
\begin{array}{|c|c|c|c|} \hline
\OO_{11} & \nbmu-2\nmu+2r &
\OO_{12} & n_{LNV}-n_\mu+r   \\\hline
\OO_{13} & n_{LNV}-\nmu+r &
\OO_{14} & n_{BNV}-\nmu +r \\\hline
\OO_{15} & \nbmu+\nmu-r  \\\cline{1-2}
\end{array}
\]
\end{center}
\label{d5Ntable}
\caption{Horizontal charges $\HH[\OO_i]$ of the leading component of the operator $\kappa_i \OO_i$ in eq. \eqref{d5Nop}. }
\end{table}
Their horizontal charges are given in table \thetable. 
For fractional $\nbmu$ and $n_{LNV}$, the only operator left is $\OO_{14}=\frac1{M_P}\bar u_i\bar d_j\bar d_k N$: its coefficients have the same structure as $\lambda_{ijk}''$, and the overall magnitude differs by a factor $\frac{\vev N}{M_P}\eps^{\nmu}$; its contribution to baryon number violation is then negligible.

\section{$SU(3)^5$ embedding of the horizontal symmetry}
\label{mfv}

In the limit that the Higgs Yukawa couplings and soft terms are  vanishing, the $R$-parity conserving MSSM is invariant under a $U(3)^5$ flavor symmetry group, under which each superfield transforms independently from the others:\footnote{
Because the fields $L_i$ and $\phi_d$ have the same gauge quantum numbers, the $R$-parity violating MSSM is actually invariant under the flavor group $U(3)^4\times U(4)\times U(1)_{\phi_d} $. An analysis of MFV supersymmetry with this symmetry group is performed in \cite{Arcadi:2011ug}. In this section, we will  take the horizontal symmetry as a diagonal subgroup of $SU(3)^5$, in order to compare our approach to the MFV approaches based on $U(3)^5$ \cite{Nikolidakis:2007fc,Csaki:2011ge}.}
\beq
U(3)^5=U(3)_Q\times U(3)_{\bar d}\times U(3)_{\bar u}\times U(3)_L\times U(3)_{\bar \ell}\ .
\eeq
The minimal flavor violation (MFV) hypothesis \cite{Nikolidakis:2007fc} assumes that the only flavor violation in BSM physics comes from the Yukawa couplings, which are treated as spurions of the $SU(3)^5$ global symmetry. It has been shown that this automatically suppresses the \rps\, couplings and can give an interesting phenomenology for natural SUSY, which evades current LHC bounds for light superpartners \cite{Csaki:2011ge}. Although possible, this is a stringent hypothesis, and by assuming that all the flavor physics is determined by the already known Yukawa parameters, it gives more weight to couplings whose discovery might just have been an historical accident.

In this section, we will investigate a weaker hypothesis, in which the horizontal symmetry is embedded in the abelian part of the $U(3)^5$ flavor group. Because each $U(3)$ has three diagonal generators, by taking a linear combination it is possible to give independent charges to fields in different generations, thus reproducing any horizontal symmetry. Instead, we will consider the embedding of the horizontal symmetry in the subgroup $SU(3)^5$.

For each $U(3)_k$, we can write $U(3)_k=SU(3)_k\times U(1)_k$, and from each $SU(3)_k$ we can extract two Abelian generators, $ T_3=\diag(1,-1,0), \ T_8=\diag(1,1,-2)$.
Collecting all the Abelian factors under which the Lagrangian is invariant gives
\beq\label{abel}
\prod_{k=Q,\bar d,\bar u,L,\bar \ell}  (T_3^k \times T_8^k)\times U(1)_Y\times U(1)_B \times U(1)_L \times U(1)_X\,,
\eeq
where we can take linear combinations the five $U(1)_k$'s to get the hypercharge $U(1)_Y$ (with charges $\frac16,\frac13,-\frac23,-\frac12,1,\frac12,-\frac12$ for $Q,\,\bar d,\,\bar u,\,L,\, \bar\ell,\phi_u,\phi_d$ respectively), the baryon number $U(1)_B$ (with charges $\frac13,-\frac13,-\frac13$ for $Q,\bar d,\bar u$), the lepton number $U(1)_L$ (with charges $+1,-1$ for $L,\bar\ell$), and the PQ  symmetry $U(1)_X$ under which $\phi_d$ has charge $-1$ and $\bar d$ and $\bar \ell$  have charge $+1$.

Of these terms, the first two give different charges to fields in different generations, while the last four are diagonal in generation space. As a transformation under the \uone's will not change the hierarchies, it is possible to see if they can be generated by the $SU(3)^5$ group alone. Thus, we write
\beq
\HH=\HH'\times \HH_{diag} \,\qquad \HH'=\prod_{k=Q,\bar d,\bar u,L,\bar\ell}  T_3^k \times T_8^k\,.
\eeq
We are looking for charges under $\HH'$ such that the charges differences $\Phi_{ij}'$ are integers; this corresponds to the horizontal symmetry being a subgroup of $SU(3)^5$.

Because $SU(3)$ is traceless, the sum of the $\HH'$ charges of the same field over the three generation has to be zero:
\beq
\HH'[k_1]+\HH'[k_2]+\HH'[k_3]=0, \qquad k=Q,\bar d,\bar u,L,\bar\ell
\eeq
In the following, we denote the charge of the field $\Phi$ under $\HH'$ as $\Phi'$; we have $\Phi_{ij}=\Phi'_{ij}$ and $\phi_a+Q_i+a_i=Q_i'+a_i',\ a=d,u$ (and an analogous expression for the leptons): as before, the hierarchies \eqref{hier} imply the solutions \eqref{diff} for the charge differences, with the constraints
\beq
Q'_3+u'_3=r, \quad Q'_3+d'_3=2-x_\beta+r ,\quad L'_3+\ell'_3=2-x_\beta+r
\eeq
In this scenario, we have five additional  constraints to satisfy, corresponding to the tracelessness of \sut. In particular, one can satisfy $\sum_i Q'_i=0$ or $\sum_i u_i'=0$, but not both at the same time (the same happens for $d,L,\ell$): we have
\beq
0=\sum_i (Q'_i+u'_i)=12+3r
\eeq
This can be solved in two ways: 
\begin{itemize}
\item using a discrete horizontal symmetry, $Z_{N}, \ N=12+3r$. Although the relation for the down quarks is similar, $\sum_i (Q_i'+d_i')=12+3r-3x_\beta$, the index $N$ is the same only for $\tan\beta\sim 3$, while for the leptons (keeping $x_\beta=0$) we need to satisfy $\sum_i (L_i'+\ell_i')=15+3r$ and a second horizontal symmetry $Z_M,\ M=15+3r$ would be needed.
\item finding a way to account for a prefactor of $\eps^4$ in the  Yukawa couplings for $\bar u_i$; then the hierarchy $(m_t/v,m_c/v,m_u/v)=(1,\eps^4,\eps^8)$ would be $(\eps^{-4},\eps^0,\eps^4)$, which can be embedded in a traceless \sut. 
\end{itemize}
We will investigate the second possibility: to be more specific, let us assume we can write the Yukawa couplings as
\beq
 \eps^M \eps^{m_{ij}} \phi_dQ_i\bar d_j+\eps^N\eps^{n_{ij}} \phi_uQ_i\bar u_j+\eps^P\eps^{p_{ij}}\phi_dL_i\bar\ell_j
\eeq
where the $m_{ij},n_{ij},p_{ij}$'s are explained by $\HH'$ and the prefactors $\eps^M,\eps^N,\eps^P$ would be explained by an additional symmetry; we will return on this aspect at the end of the section. $M,N,P$ are determined by requiring  $\sum_i(Q_i'+u_i')=0, \ \sum_i(Q_i'+d_i')=0,\ \sum_i(L_i'+\ell_i')=0$:
\beq
N-r=4,\qquad M-r=4-x_\beta,\qquad P-r=5-x_\beta
\eeq
\begin{table}[bdp]
\begin{center}
$
\begin{array}{c|ccc|ccc|ccc|ccc|ccc|cc}\
\Phi&Q'_1&Q'_2&Q'_3&d'_1&d'_2&d'_3&u'_1&u'_2&u'_3&L'_1&L'_2&L'_3&\ell'_1&\ell'_2&\ell'_3&\phi_u&\phi_d\\\hline
\HH'&\frac43& \frac13& -\frac53& \frac23&- \frac13& -\frac13& \frac83& -\frac13& -\frac73&0&0&0&3&-1&-2&0&0\\\hline
\HH_{diag}&\multicolumn{3}{c|}{-b-y}&\multicolumn{3}{c|}{b-2y+x}&\multicolumn{3}{c|}{b+4y}&\multicolumn{3}{c|}{l+3y}&\multicolumn{3}{c|}{-l-6y+x}&-3y&3y-x\\
\end{array}$
\end{center}
\label{tablemfv}
\caption{The charges of the fields under $\HH'$, the non-diagonal part of the horizontal symmetry \eqref{hh'}, and the diagonal contribution if the full symmetry is $\HH=\HH'-3bB-6yY+xX+lL$, with the normalizations and signs chosen to have simple coefficients.}
\end{table}
With this addition, all the non-diagonal charges are uniquely determined in table \thetable \ and the horizontal symmetry can be written as a linear combination of all the Abelian generators.
%
%
\beq\label{hh'}
\HH'=\frac12T_3^Q+\frac56T_8^Q+\frac12T_3^{\bar{d}}+\frac16T_8^{\bar{d}}+\frac32T_3^{\bar{u}}+\frac76T_8^{\bar{u}}+2T_3^{\bar\ell}+T_8^{\bar\ell}
\eeq

These charges give the same charge differences as in \eqref{diff}, except for the leptons, $(\ell_{12},\ell_{13},\ell_{23})=(4,5,1)$ instead of $(4,6,2)$ (here we assume that some of the $\OO(1)$ factors in front of the superpotential can lead to a slightly different hierarchy, with $m_\mu/m_\tau=\eps$ instead of $\eps^{1.9}$: e.g., we can have couplings of the form  $0.4\phi_dL_3\ell_3 +1.5\phi_dL_2\ell_2$).
 Then, the RPV coupling textures  in section \ref{rpv} are the same, apart from the structure of the operator $\lambda_{ijk}L_iL_j\bar\ell_k$, whose couplings have  rows that now read $(\eps^5,\eps,1)$ instead of $(\eps^6,\eps^2,1)$. In particular, to have a phenomenologically viable model, the LNV couplings are still forbidden, and this can be done by having a fractional value for $\nbmu$. The experimental limits on $n_{BNV}$ are the same as above.

\subsection*{The prefactors}
In the last section we have seen how the relative hierarchies between the quarks and leptons can be understood in terms of a horizontal symmetry that is a subgroup of the flavor group $SU(3)^5$. We did not discuss the origin of the absolute scale, which appears as an overall factor of $\eps^4$ in front of the quarks Yukawas and as a factor of $\eps^5$ for the leptons. 

We can easily get that factor by adding an extra horizontal $U(1)'$ and having the horizontal symmetry to be $U(1)'\times \HH$. The charges of the fields under $U(1)'$ can just be arranged to get the required suppression, e.g. by taking the charges of $(\phi_u,\phi_d,Q,\bar d,\bar u,L,\bar \ell)$ to be $(0,0,2,2,2,2,3)$.

Another option would be to have a loop factor: if all the SM Yukawas are generated through a loop diagram, it is natural for them to have a factor of $\frac{g^2}{16\pi^2}\sim\eps^{4-5}$.

These considerations do not change the textures of the RPV couplings in section \ref{rpv}, which are determined by the charge differences in eq. \eqref{diff}.

\section{Conclusions}
\label{conclusions}
Motivated by current LHC searches that  present no hint for $R$-parity conserving supersymmetry, we have revisited \rps \ models where a horizontal symmetry is responsible for the hierarchies of the SM fermion mass  and mixing hierarchies. In this case, the LSP can decay in the detector, thus leaving no large missing energy events.
The RPV couplings are hierarchical and their textures are completely fixed by the known hierarchies for squarks and leptons. In particular, there is no need to impose anomaly cancellation through a standard Green-Schwartz mechanism. The phenomenology is similar to the MFV SUSY scenario, with the largest RPV coupling involving the stop. 
While the overall scale of the RPV operators is not fixed by the horizontal symmetry, null LHC searches generically forbid light superpartners with \rps \ coefficients smaller than about $10^{-9}$.
With this bound, if lepton and baryon number are violated at the same time the proton lifetime would be shorter than its current experimental limit. We are then led to consider just baryon number violation, while lepton number is conserved in the renormalizable superpotential. The largest $\bar u\bar d \bar d$ coupling, $\lambda''_{323}=\eps^{n_{BNV}}$,  has to lay between $10^{-3}$ and $10^{-9}$, corresponding to $4\lesssim n_{BNV}\lesssim13$. In particular, we stress that for $8<n_{BNV}<13$ (corresponding to $10^{-9}<\lambda''< 10^{-6}$, about half of the remaining allowed range for $\lambda''$), displaced vertices would be a striking signature of hadronic stop decays. 
As the LHC pushes up the limits on R-hadrons, light LSPs with arbitrary small $R$-parity violating coefficients will be excluded to higher and higher squark masses, thus strengthening our argument for natural SUSY with $B$ violation.



\appendix*
\section{The neutrino sector}
Let us recall the neutrino oscillation parameters:
\beq\label{Dms}
\Delta m_\odot^2=7.45\cdot10^{-5}\text{ eV}^2, \qquad |\Delta m_A|^2=2.35\cdot10^{-3}\text{ eV}^2
\eeq
We take note of  the relation $\Delta m_\odot^2 \sim \eps^2  |\Delta m_A|^2$. We show two ways of generating neutrino masses:

\begin{itemize}
\item Dirac neutrino masses: if there are three right handed neutrinos with a mass term ${g^\nu}_{ij} L_i\phi_u\bar \nu_{Rj}$, the neutrino have Dirac masses of order $g^\nu v$. 
The oscillation data can be fitted by $g^\nu_{11}\sim g^\nu_{22}=3.5\cdot 10^{-14}\approx \eps^{21}$ and $g^\nu_{33}\sim 2\cdot10^{-13}\sim\eps^{20}$. The horizontal symmetry relates the magnitude of the couplings to the charges of the fields:
\beq
(g^\nu)_{ij}=\eps^{L_i+\phi_u+\nu_j} \sim\eps^{\nbmu +L_{i3}+\nu_j}
\eeq
As the lepton sector is similar to the quark sector, the expression \eqref{lept} for the PMNS matrix is analogous to the expression for the CKM matrix.
Taking $\nu_i=\frac{15}2,\ L_{i3}=(1,1,0)$ and $\nbmu=\frac{25}2$, the atmospheric and solar oscillation parameters are explained by a set of charges of order 7-8, which are deemed natural. The PMNS matrix has the  third form  in eq. \eqref{pmns}. Other natural choices of charges are possible to have a non-hierarchical mixing matrix.

\item type I seesaw mechanism: here we mainly review arguments given in \cite{Sato:1997hv,Dreiner:2003yr}.
Including three heavy right handed neutrinos $N_R$ with a superpotential
\beq
g_\nu^{ij}L_i\phi_u N_{R j}+\frac12 \bar N_R^c M_{RR} N_R,
\eeq
the magnitude of the couplings $g_\nu$'s are
\beq
g_\nu^{ij}=\eps^{L_i+\phi_u+N_j}
\eeq
with the definition $N_j=\HH[N_{Rj}]$, and  $(M_{RR})_{ij}=M_R\eps^{N_i+N_j}$. Assuming $N_i>0$, the seesaw mechanism gives Majorana masses for the neutrinos as
\beq
m^{ij}_{\nu}=\frac{v^2}{2{M_R}}\eps^{2\phi_u+L_i+L_j}=\frac{v^2}{2{M_R}}\eps^{2\nbmu}\eps^{L_{i3}+L_{j3}},
\eeq
where $\nbmu=L_3+\phi_u-r$ is defined in eq. \eqref{2lnv}. For $M_R=M_P$, the masses are too small, although it could be $M_R=M_{GUT}$, or $M_R=M$, the flavor breaking scale.\footnote{The discussion for $N_i<0$ or $\nbmu<0$ can be found in \cite{Dreiner:2003yr}; the overall scale $v^2/2M_R$ can be enhanced and be consistent with the experimental results \eqref{Dms} even for $M_R=M_P$.}

Given the expression above for $m^{ij}_{\nu}$, the PMNS matrix $U$ still take the form $|U_{ij}|=\eps^{|L_i-L_j|}$. If we take the PMNS matrix to be anarchical, that is, $|U_{ij}|=\OO(1)$ for all $i,j$, we have $L_{i3}=0$, and all masses should be of order $\frac{v^2}{2{M_R}}$. To explain oscillation data, we have to assume a tuning of order $\eps$ in the eigenvalues of the matrix of $\OO(1)$ coefficients, e.g.  by having eigenvalues such as
\beq
m_{\nu j}=\frac{v^2}{2{M_R}} (1,1,\eps^{-1})
\eeq
For a seesaw mechanism, some amount of tuning is generally needed. The masses are
\beq
m_\nu=\frac{v^2}{2{M_R}}  \eps^{\left(\begin{array}{ccc} 2L_{13} & L_{13}+L_{23}  & L_{13}  \\ L_{23}+L_{13} & 2L_{23}  & L_{23}  \\ L_{13} & L_{23}  & 0 \end{array}\right)}.
\eeq
As $L_{i3}$ is an integer, the eigenvalues will either be degenerate or differ by a factor of at least $\eps^2$; the latter case would corresponds to a factor of $\eps^4$ between the atmospheric and solar squared mass differences, instead of $\eps^2$. Thus, an accidental tuning of the $\OO(1)$ factors is needed to have the mass eigenvalues differ by a factor of $\eps$; for clarity, taking $L_{i3}=(1,1,0)$, we could have
\beq
m_{\nu j}=\frac{v^2}{2{M_R}}   (A \eps^2, B\eps^2, C^2)
\eeq
which, for $A=\frac12, B=\frac23$ and $ C=2$, gives $\Delta m_\odot^2 \sim \eps^2  |\Delta m_A|^2$.

\end{itemize}

\section{The other solution for $Q_{ij}$}
The second solution in eq. \eqref{diff} is
\beq\label{diff2}
\begin{array}{|ccc|ccc|ccc|ccc|ccc|}\hline
Q_{12}&Q_{13}&Q_{23}&d_{12}&d_{13}&d_{23}&u_{12}&u_{13}&u_{23}&L_{12}&L_{13}&L_{23}&\ell_{12}&\ell_{13}&\ell_{23}\\\hline
-1&-3&-2&  3&8&5&  5&12&7 &0&0&0&4&6&2\\\hline
\end{array}
\eeq

Because the charge differences in the baryon sector have changed, the \rps \ phenomenology  will change too. Factoring out the dependence on $n_{LNV}$ and $n_{BNV}$ gives the following textures, which can be compared to eqs. \eqref{2lnv}--\eqref{udd}:
\beq
&&\bar\mu_1=\bar\mu_2=\bar\mu_3=m\eps^{n_{\bar\mu}},\qquad
\left(\begin{array}{ccc} \lambda_{121}  & \lambda_{122}  & \lambda_{123}  \\ \lambda_{131} & \lambda_{132}  &\lambda_{133}   \\\lambda_{231}  & \lambda_{232}  & \lambda_{233} \end{array}\right)=\eps^{n_{LNV}}\left(\begin{array}{ccc}\eps^6 & \eps^2 & 1 \\ \eps^6 & \eps^2 & 1 \\ \eps^6 & \eps^2 & 1\end{array}\right),
\\\nn
&&\left(\begin{array}{ccc} \lambda'_{i11}  & \lambda'_{i12}  & \lambda'_{i13}  \\ \lambda'_{i21} & \lambda'_{i22}  &\lambda'_{i23}   \\\lambda'_{i31}  & \lambda'_{i32}  & \lambda'_{i33} \end{array}\right)=\eps^{n_{LNV}}\left(\begin{array}{ccc}\eps^5 & \eps^2 & \eps^{-3} \\ \eps^6 & \eps^3 & \eps^{-2} \\ \eps^8 & \eps^5 & 1\end{array}\right)\,,
\left(\begin{array}{ccc} \lambda''_{112}  & \lambda''_{212}  & \lambda''_{312}  \\ \lambda''_{113} & \lambda''_{213}  &\lambda''_{313}   \\\lambda''_{123}  & \lambda''_{223}  & \lambda''_{323} \end{array}\right)=\eps^{n_{BNV}}\left(\begin{array}{ccc}\eps^{20} & \eps^{15} & \eps^{8} \\ \eps^{15} & \eps^{10} & \eps^{3} \\ \eps^{12} & \eps^7 & 1\end{array}\right)
\eeq
where $\nbmu, n_{LNV}$ and $n_{BNV}$ have been defined as in the main text. This gives a different phenomenology if we had lepton number violation, in which the largest $\slashed L$ coupling is $L_iQ_{1}\bar d_3$. For BNV, the phenomenology is approximately the same as before, with a dominant coupling $\lambda_{323}''$, except that there is more suppression on couplings involving the  first 2 generations.

As before, LHC searches for stable particles or missing energy events would apply if the couplings were small enough: the  bound is still $n_{LNV},n_{BNV}\lesssim 13$. The constraint on proton decay used earlier (from $p\to K^+\nu$) is weaker in this case, because the first generations of squarks have smaller couplings
\beq
|\lambda'_{i23}\lambda''_{113}|=\eps^{n_{LNV}+n_{BNV}+13}\lesssim10^{-{27}} ( m_{\tilde b_{R}}/100)^2=\eps^{41}( m_{\tilde s_{R}}/100)^2
\eeq
which gives the constraint $n_{BNV}+n_{LNV}\gtrsim28$. Other decay channels give comparable bounds:
\beq
p\to \pi^0\ell^+:&\ &|\lambda'_{l13}{\lambda''}^*_{113}|=\eps^{n_{LNV}+n_{BNV}+12}\lesssim 10^{-25}\left(\frac{m_{\tilde b_{R}}}{500\text{ GeV}}\right)^2, \qquad n_{LNV}+n_{BNV}\gtrsim 27.\nn\\\nn
p\to K^0\ell^+:&\ &|\lambda'_{l13}\lambda''_{123}|=\eps^{n_{LNV}+n_{BNV}+9}\lesssim 10^{-25}\left(\frac{m_{\tilde b_{R}}}{500\text{ GeV}}\right)^2,\qquad n_{LNV}+n_{BNV}\gtrsim 30.
\eeq
These bounds are incompatible with the LHC limit $n_{LNV},n_{BNV}\lesssim 13$ required by not having light supersymmetric particles that are stables on collider timescales. Thus we will set to zero all the LNV couplings by taking $n_{LNV}$ (and $\nbmu$) fractional, and we are left with the BNV operator $\bar u\bar d \bar d$. Revisiting the limits on $\lambda''$  from low energy experiments, eqs. \eqref{nn}--\eqref{bb}, we get a lower limit on $n_{BNV}\gtrsim 3$ from $n-\bar n$ oscillation; the allowed range for $n_{BNV}$ is approximately the same as for the other solution allowed by  horizontal symmetries. The only phenomenological difference is that the hierarchy between $|\lambda_{323}''|$ and the other coefficients is enhanced, and $|\lambda_{323}''|$ is the only coupling that could realistically be measured.


\begin{acknowledgments}
The author would like to thank Michael Dine for proposing the topic studied and for providing ongoing support and discussion, as well as for feedback on the manuscript of this work.
\end{acknowledgments}


%

\end{document}